\documentstyle[epsf]{mn}

\newif\ifAMStwofonts

\ifoldfss
  \ifCUPmtlplainloaded \else
    \NewTextAlphabet{textbfit} {cmbxti10} {}
    \NewTextAlphabet{textbfss} {cmssbx10} {}
    \NewMathAlphabet{mathbfit} {cmbxti10} {} 
    \NewMathAlphabet{mathbfss} {cmssbx10} {} 
  \fi
  \ifAMStwofonts
    \ifCUPmtlplainloaded \else
      \NewSymbolFont{upmath} {eurm10}
      \NewSymbolFont{AMSa} {msam10}
      \NewMathSymbol{\upi}     {0}{upmath}{19}
      \NewMathSymbol{\umu}     {0}{upmath}{16}
      \NewMathSymbol{\upartial}{0}{upmath}{40}
      \NewMathSymbol{\leqslant}{3}{AMSa}{36}
      \NewMathSymbol{\geqslant}{3}{AMSa}{3E}

    \fi
  \fi
\fi 

\ifnfssone
  \newmathalphabet{\mathit}
  \addtoversion{normal}{\mathit}{cmr}{m}{it}
  \addtoversion{bold}{\mathit}{cmr}{bx}{it}
  \newmathalphabet{\mathbfit} 
  \addtoversion{normal}{\mathbfit}{cmr}{bx}{it}
  \addtoversion{bold}{\mathbfit}{cmr}{bx}{it}
  \newmathalphabet{\mathbfss} 
  \addtoversion{normal}{\mathbfss}{cmss}{bx}{n}
  \addtoversion{bold}{\mathbfss}{cmss}{bx}{n}
  \ifAMStwofonts
    \ifCUPmtlplainloaded \else
      \UseAMStwoboldmath
      \makeatletter
      \new@mathgroup\upmath@group
      \define@mathgroup\mv@normal\upmath@group{eur}{m}{n}
      \define@mathgroup\mv@bold\upmath@group{eur}{b}{n}
      \edef\UPM{\hexnumber\upmath@group}
      \new@mathgroup\amsa@group
      \define@mathgroup\mv@normal\amsa@group{msa}{m}{n}
      \define@mathgroup\mv@bold\amsa@group{msa}{m}{n}
      \edef\AMSa{\hexnumber\amsa@group}
      \makeatother
      \mathchardef\upi="0\UPM19
      \mathchardef\umu="0\UPM16
      \mathchardef\upartial="0\UPM40
      \mathchardef\leqslant="3\AMSa36
      \mathchardef\geqslant="3\AMSa3E
    \fi
  \fi
\fi 

\ifnfsstwo
  \DeclareMathAlphabet{\mathbfit}{OT1}{cmr}{bx}{it}
  \SetMathAlphabet\mathbfit{bold}{OT1}{cmr}{bx}{it}
  \DeclareMathAlphabet{\mathbfss}{OT1}{cmss}{bx}{n}
  \SetMathAlphabet\mathbfss{bold}{OT1}{cmss}{bx}{n}
  \ifAMStwofonts
    \ifCUPmtlplainloaded \else
      \DeclareSymbolFont{UPM}{U}{eur}{m}{n}
      \SetSymbolFont{UPM}{bold}{U}{eur}{b}{n}
      \DeclareSymbolFont{AMSa}{U}{msa}{m}{n}
      \DeclareMathSymbol{\upi}{0}{UPM}{"19}
      \DeclareMathSymbol{\umu}{0}{UPM}{"16}
      \DeclareMathSymbol{\upartial}{0}{UPM}{"40}
      \DeclareMathSymbol{\leqslant}{3}{AMSa}{"36}
      \DeclareMathSymbol{\geqslant}{3}{AMSa}{"3E}
    \fi
  \fi
\fi 

\ifCUPmtlplainloaded \else
  \ifAMStwofonts \else 
    \def\upi{\pi}
    \def\umu{\mu}
    \def\upartial{\partial}
  \fi
\fi

\title{An Extremely Slow Nova?}
\author[Patricia Whitelock \& Freddy Marang]
       {Patricia Whitelock\footnote{email: paw@saao.ac.za} \& Freddy Marang \\
SAAO, PO Box 9, Observatory, 7935, South Africa (email: paw@saao.ac.za). }
\date{\today }
\pagerange{\pageref{firstpage}--\pageref{lastpage}}
\pubyear{2001}

\begin{document}

\maketitle

\label{firstpage}

\begin{abstract}
 The binary companion to the peculiar F supergiant HD\,172481 is shown to be
a Mira variable with a pulsation period of 312 days. Its characteristics are
within the normal range found for solitary Miras of that period, although
its pulsation amplitude and mass-loss rate $\dot{M} \sim 3 \times 10^{-6}\,
M_{\odot}yr^{-1}$ are higher than average. Reasons are given for suspecting
that the F supergiant, which has $L\sim 10^4\,L_{\odot}$, is a white dwarf
burning hydrogen accreted from its companion.

\end{abstract}


\section{Introduction}

HD\,172481 is a high galactic latitude F-type supergiant which has drawn
some attention since the IRAS survey showed it to have an infrared dust
excess (Odenwald 1986).  Reyniers \& Van Winckel (2000, henceforth RVW) have
since shown that the star is a spectroscopic binary with a cool M-giant
companion. Their detailed spectroscopic analysis of the F supergiant
indicates a moderate metal deficiency ([Fe/H]=--0.55), no CNO-enhancement, a
slight enhancement of s-process elements and high lithium content ($\log
\epsilon(\rm Li)=3.6$). Essentially nothing was known about the M star prior
to the present paper.

Since the IRAS measurements most authors have assumed that the F supergiant
is a post-AGB star (RVW and references therein). This paper first discussed
the nature of the M giant (section 3) and then, more speculatively, provides
a possible explanation for the F supergiant (section~4).

\section{Photometry}
 The published Stromgren and Geneva photometry (Houk \& Mermilliod 1998)
indicate $V$ between 8.9 and 9.1 while Geneva photometry from RVW shows that
the visual magnitude is variable with a peak to peak amplitude of 0.22 mag. 
RVW show that the M star contributes about 10 percent of the flux at $V$
when the Mira is at maximum light (see below). The value, $V=9.09$ mag,
given in the {\sc simbad} database is used in the following analysis as
representative of the mean light of the F star.

 Forty-five $JHKL$ measurements were made between 1988 and 1999 using the
MkII photometer on the 0.75-m telescope at SAAO Sutherland. These are
on the SAAO system (Carter 1990) and are listed in Table 1. They are 
accurate to about $\pm 0.03$\,mag at $JHK$ and $\pm 0.05$\, mag at $L$. 

\begin{table}
\centering
\begin{minipage}{140mm}
\caption{Observed  Magnitudes}
\begin{tabular}{@{}rcccc@{}}
JD &   $J$  &  $H$ & $K$ & $L$ \\
$-2440000$&\multicolumn{4}{c}{(mag)}\\
7252.63 & 6.693	& 5.957	& 5.644	& 5.197\\
7270.65	& 6.671	& 5.918	& 5.606	& 5.138\\
7328.54	& 6.740	& 5.967	& 5.611	& 5.166\\
7334.49	& 6.742	& 5.997	& 5.639	& 5.175\\
7356.41	& 6.874	& 6.156	& 5.775	& 5.303\\
7362.43	& 6.934	& 6.224	& 5.821	& 5.422\\
7372.43	& 7.005	& 6.317	& 5.898	& 5.444\\
7383.42	& 7.059	& 6.384	& 5.965	& 5.510\\
7393.35	& 7.187	& 6.551	& 6.111	& 5.606\\
7422.34	& 7.196	& 6.544	& 6.094	& 5.553\\
7461.30	& 7.097	& 6.436	& 5.996 &      \\  
7617.64	& 6.687	& 5.878	& 5.499	& 4.922\\
7685.52	& 7.069	& 6.388	& 5.940	& 5.376\\
7698.48	& 7.106	& 6.437	& 6.002	& 5.357\\
7709.45	& 7.141	& 6.494	& 6.050	& 5.507\\
7719.46	& 7.167	& 6.518	& 6.092	& 5.557\\
7737.37	& 7.102	& 6.473	& 6.042	& 5.507\\
7742.45	& 7.128	& 6.477	& 6.050	& 5.541\\
7760.33	& 7.111	& 6.455	& 6.027	& 5.481\\
7767.31	& 7.095	& 6.437	& 6.020	& 5.459\\
7816.24	& 7.030	& 6.386	& 5.977	& 5.327\\
7823.28	& 6.971	& 6.313	& 5.926	& 5.291\\
7836.26	& 6.904	& 6.192	& 5.807	& 5.170\\
8054.57	& 7.211	& 6.580	& 6.133	& 5.559\\
8066.56	& 7.204	& 6.554	& 6.111	& 5.614\\
8093.53	& 7.139	& 6.479	& 6.050	& 5.437\\
8115.43	& 7.149	& 6.495	& 6.073	& 5.488\\
8166.35	& 6.856	& 6.130	& 5.751	& 5.177\\
8172.35	& 6.788	& 6.057	& 5.703	& 5.165\\
8451.48	& 7.017	& 6.341	& 5.937	& 5.336\\
8453.48	& 7.006	& 6.326	& 5.919	& 5.372\\
8853.36	& 6.617	& 5.820	& 5.461	& 4.953\\
9146.65	& 6.745	& 5.948	& 5.595	& 5.134\\
9223.42	& 6.876	& 6.174	& 5.793	& 5.276\\
9583.44	& 7.195	& 6.438	& 6.033	& 5.709\\
9615.39	& 7.229	& 6.589	& 6.167	& 5.568\\
9641.32	& 7.219	& 6.584	& 6.135	& 5.650\\
10238.57& 7.251	& 6.615	& 6.201	& 5.691\\
10589.61& 7.160	& 6.477	& 6.068	& 5.579\\
10716.35& 6.607	& 5.795	& 5.429	& 4.915\\
10755.25& 6.723	& 5.924	& 5.553	& 5.071\\
11100.27& 6.965	& 6.239	& 5.808	& 5.398\\
11299.61& 6.754	& 5.986	& 5.645	& 5.173\\
11388.40& 6.788	& 6.068	& 5.684	& 5.202\\
11675.55& 6.758 & 5.983 & 5.584 & 5.163\\
11711.57& 6.904 & 6.178 & 5.759 & 5.222\\
\end{tabular} 
\end{minipage}
\end{table}

\section{Nature of the Cool Star}
 A Fourier analysis of the data in Table 1 reveals a clear period of 312
days and light curves folded on this period are shown in Fig. 1.  The peak
to peak amplitudes of the Fourier fits to these variations are: $\Delta J =
0.53$, $\Delta H = 0.68$, $\Delta K = 0.62$ and $\Delta L = 0.58$ mag. These
of course are lower limits to the amplitudes of the M giant as the F
supergiant will contribute to the observed magnitude, particularly at the
shorter wavelengths. Corrected amplitudes are discussed in section 3.1 
and listed in Table~2.

\begin{figure}
\centering
\epsfxsize=8.4cm
\epsffile{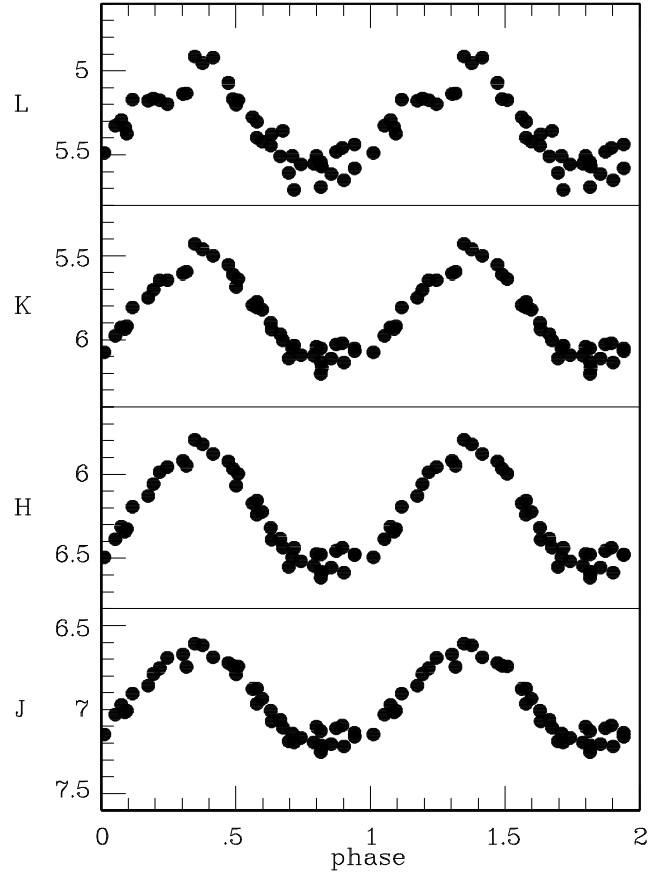}
\caption{The infrared observations for HD\,172481 phased on a period of 312
days, with an arbitrary zero-point of JD\,2440000; each point is plotted
twice.}
\end{figure}

 These large amplitude variations clearly demonstrate that the M giant is
a Mira variable and hence at the very tip of the Asymptotic Giant Branch
(Feast \& Whitelock 1987).

\subsection{Infrared Colours}
 The reddening from HD\,172481 has been estimated as $E(B-V)=0.44$ (RVW) and
0.27 (Bersier 1996). In the following we calculate the magnitudes and
colours using both of these values, generally giving the value for
$E(B-V)=0.44$ first and following it with the value associated with
$E(B-V)=0.27$, in brackets.

\begin{table}
\centering
\begin{minipage}{140mm}
\caption{Mean Magnitudes}
\begin{tabular}{@{}lccccc@{}}
component &   $V$  &  $J$ & $H$ & $K$ & $L$ \\
&\multicolumn{5}{c}{(mag)}\\
max &  -   & 6.65 & 5.85 & 5.49 & 4.99 \\
min &  (9.09) & 7.18 & 6.53 & 6.11 & 5.57 \\
\multicolumn{5}{l}{$ -E{(B-V)}=0.44$} \\
F2Ia & 7.75 & 7.21 & 7.04 & 7.00 & 6.95 \\
Mira (mean) &  -   & 7.58 & 6.58 & 6.08 & 5.47 \\
$\Delta$ & -  & 1.3  & 1.1  & 0.9  & 0.7 \\
\multicolumn{5}{l}{$ -E{(B-V)}=0.27$} \\
F2Ia & 8.27 & 7.73 & 7.56 & 7.52 & 7.47 \\
Mira (mean)&   -  & 7.29 & 6.41 & 5.98 & 5.39 \\
$\Delta$ & -  & 0.9 & 0.9 & 0.8 & 0.7\\
\end{tabular}
\end{minipage}
\end{table}

 The first two lines of Table 2 list the magnitudes for HD\,172481 at the
Fourier maximum and minimum (note that, as seen in Fig.~1, the phase shifts
between the $JHKL$ light-curves are negligible).  The next three lines have
been corrected assuming a reddening of $ E{(B-V)}=0.44$ and give,
respectively, the reddened corrected magnitudes of the F2Ia star (from
Johnson 1966) on the assumption that this is the only significant
contributor at $V$, the reddening corrected mean magnitude of the Mira after
the contribution from the F star has been removed, and the peak to peak
amplitude of the Mira variations, also after removing the contribution of
the F star. These three lines are then repeated using the alternative,
$E{(B-V)}=0.27$, reddening correction.

 It is instructive to compare the colours deduced for this Mira with those
of normal isolated Miras. Whitelock, Marang \& Feast (2000, henceforth WMF)
provide $JHKL$ observations of Miras with thin dust shells, while Whitelock
et al. (1994) do the same for IRAS selected Miras with relatively thick
shells. The mean colours of the Mira are listed in Table 3 together with
those of a typical 312 day Mira, calculated from equations 1 to 4 of WMF.

 The deduced colours are obviously somewhat uncertain, particularly at $J$,
because of the uncertain correction for the F star.  Nevertheless, they fall
within the range shown by normal unreddened Miras. In particular, $K-L$,
although larger than the mean calculated from WMF equation 3, is within the
range shown by normal Miras. WMF note a strong correlation between the
pulsation amplitude and $K-L$, and indeed the pulsation amplitude of this
Mira is high.

 This Mira is of particular interest in view of the fact that we can infer
its metallicity; only a very small number of Miras, those associated with
globular clusters, have reliable metallicities. RVW estimated [Fe/H]=--0.55
for the F supergiant which we might reasonably assume also applies to the
Mira. For comparison, Table 3 gives the colours (from WMF) of the Mira V3 in
NGC\,5927 which has a 311 day period and [Fe/H]=--0.37 (Harris 1996).

\subsection{Mass-loss Rate}
 As RVW discussed, HD\,172481 was detected by IRAS and shows an obvious
infrared excess. The colour corrected 12\,$\mu$m mag is [12]=1.76, so
$K-[12]=4.3$ (4.2). This is larger than values shown by any of the stars
discussed by WMF, but is within the range found for IRAS Miras by Whitelock
et al. (1994). This $K-[12]$ would imply a mass-loss rate of $\dot{M} \sim 3
\times 10^{-6}\, M_{\odot}yr^{-1}$ (Whitelock et al. 1994 fig 21) which is
high for a 312 day Mira, but higher mass-loss rates are usually associated
with larger pulsation amplitudes and again we note that the pulsation
amplitude of this star is high.

 It would therefore seem possible that the dust shell, which is responsible
for large $K-[12]$ and presumably for the circumstellar extinction,
originates entirely from the Mira.  While it is not essential to assume that
any of the dust is associated with the F supergiant, we cannot rule out the
possibility that some of it is. Neither can we rule out the possibility 
of dust trapped in a circumbinary disk, as has been proposed for some other
supposed post-AGB binary systems (e.g. Waters et al. 1993; Van Winckel 
1999)

Note also that the IRAS colours of HD\,172481 are within the range found for
D-type symbiotics (Whitelock 1988), illustrating the similarity of their 
dust shells (see section~4).

\begin{table}
\centering
\begin{minipage}{140mm}
\caption{Mean Colours of various components}
\begin{tabular}{@{}lccccc@{}}
star & $J-H$  &  $H-K$  & $K-L$ & $J-K$ & $E(B-V)$\\
&\multicolumn{4}{c}{(mag)}\\
Mira & 1.00 & 0.50 & 0.61 & 1.50 & 0.44\\
Mira & 0.88 & 0.43 & 0.59 & 1.31 & 0.27\\
$P=312$ & 0.94& 0.44 & 0.51 & 1.38\\
N5927\,V3 & 0.95 & 0.39 & 0.52 & 1.34 \\
\end{tabular}
\end{minipage}
\end{table}

\subsection{Distance and Luminosity}
 The distance modulus of the Mira can be calculated via the period
luminosity relation (Whitelock \& Feast 2000):
$$ M_K=-3.47 \log P +0.84, $$
 to give $(K_0-M_K)=13.9$ (13.8) or a distance of 6.0 (5.7) kpc,
and a height above the galactic plane of --1.1 (--1.0) kpc. 

The bolometric magnitude of the Mira can be established by fitting a
blackbody to the $JHKL$ flux (see Robertson \& Feast 1981). This gives a
bolometric magnitude at mean light of 9.30 (9.09) and at maximum light of
8.72 (8.64). Or in absolute terms at mean light $M_{bol}=-4.59\,(-4.69)$ and
$L_M=5.3 \times 10^3\, (5.8 \times 10^3) L_{\odot}$.

\section{the F supergiant}
 RVW established that the relative luminosities of the F supergiant and the
M star are $L_F/L_M=1.8$, for a reddening of $E(B-V)=0.44$ (or
$L_F/L_M=1.0$ for $E(B-V)=0.2$). The near-infrared measurement which they
used to define the flux of the system was obtained on JD\,2448850 ($K=5.48$
Van Winckel, private communication), i.e. at maximum light for the Mira.
Thus the luminosity of the F star is $L_F=1.6 \times 10^4\,L_{\odot}$ ($0.9
\times 10^4\,L_{\odot}$).
 
RVW argue that the F supergiant is a post-AGB star on the basis of its high
galactic latitude, large radial velocity and moderate metal deficiency. In
support of this they also mention the dust, a variable H$\alpha$ profile
and a slight overabundance of s-process elements. Our identification of the M
star as a Mira variable confirms the binary as a low mass system. There is,
however, an alternative to the post-AGB star scenario. The F supergiant
could be a white dwarf, with hydrogen-shell burning of material accreted
from the Mira wind providing a pseudo-photosphere and the observed
luminosity.

Iben \& Tutukov (1996) discuss the evolution of mass accreting white dwarfs
in wide binaries systems, such as symbiotic stars. Of particular interest in
the present context are the D-type symbiotics where a white dwarf accretes
from the wind of a Mira variable in a binary system with a period in excess
of about 10 years (Whitelock 1987). In certain cases the accretion rates are
such that a thermonuclear runaway gives rise to a very slow nova, e.g.
RR~Tel and V1016~Cyg. These symbiotic stars, with their high excitation
emission lines, are apparently very different from HD\,172481. However, it is
clear from Iben \& Tutukov that relatively small differences, e.g. in the
accretion rate of the white dwarf, can give rise to very different phenomena.

The luminosity derived above, $L_F \sim 10^4\,L_{\odot}$, is that expected
for a cool $\sim 0.6\, M_{\odot}$ white dwarf in a hydrogen shell burning
phase following a hydrogen shell flash (see Iben \& Tutukov fig.~5).

A very rough estimate of the accretion rate of the white dwarf can be
derived from Iben \& Tutukov's equation 2. If we assume the mass-loss rate
from the Mira (section 3.2), a $0.6\, M_{\odot}$ white dwarf, a total
stellar mass (white dwarf plus Mira) of $1.5\, M_{\odot}$, a wind velocity
of $20\, \rm km\,s^{-1}$ and an orbital period of 10 years, the derived
accretion rate onto the white dwarf is $2\times 10^{-7} M_{\odot}\rm
yr^{-1}$. 

It is clear from fig.~7 of Iben \& Tutukov that this rate would in fact
result in steady hydrogen burning on the surface of the white dwarf.
However, for the parameters assumed this rate should probably be regarded as
an upper limit, as Iben \& Tutukov's equation 2 is based on Bondi-Hoyle
accretion and other estimates lead to significantly (up to ten times) lower
values (e.g. Theuns, Boffin \& Jorissen 1996; Mastrodemos \& Morris 1999).
Furthermore, plausible changes in the period or the wind velocity could
increase or decrease the rate. Nevertheless it seems likely that a white
dwarf paired with this Mira would accrete mass in such a way as to either
burn it steadily or in shell flashes, for a large range of stellar
separations.

In discussing the nova phenomenon Iben \& Tutukov point out that in wide
binaries there is no reason to expect the loss of the hydrogen envelope
after the explosion. So that it is possible for the newly created supergiant
to last for several decades. This might explain the present condition of the
F supergiant in HD\,172481.

The high abundance of lithium in the supergiant spectrum (RVW) may well be
key to understanding this system. Arnould \& N{\o}rgaard (1975) and
Starrfield et al. (1978) predict the formation of large quantities of
lithium during hydrogen-burning themonuclear runaways. However, their
calculations are not obviously applicable to the situation under discussion
and they also predict an overabundance of carbon and nitrogen which are not
found in HD\,172481 (RVW). More recently Hernanz et al. (1996) predict
significant lithium production from some nova models, but these seem to
underestimate the ejecta masses.  Clearly more theoretical work is required
in this area.

\section*{Acknowledgments}
 Our thanks to Hans Van Winckel for providing a preprint of their paper and
details of their unpublished infrared photometry. We are also grateful to
Michael Feast for helpful discussions. This paper is based on observations
made from the South African Astronomical Observatory and also made use of
information from the {\sc simbad} data-base. The comments and suggestions 
of an anonymous referee were also helpful.


\begin{thebibliography}{99}
\bibitem{} Arnould M., N{\o}rgaard H., 1975, AA, 42,55
\bibitem{} Carter B. S. C., 1990, MNRAS, 242, 1
\bibitem{} Bersier D., 1996, A\&A, 308, 514
\bibitem{} Feast M. W., Whitelock P. A., 1987, in: Late Stages of Stellar
   Evolution, (eds.) S. Kwok, S. R. Pottasch, Reidel, Dordrecht, p.~33 
\bibitem{} Harris W. E., 1996, AJ, 112, 1487 \\
     (http://physun.physics.mcmaster.ca/Globular.html)
\bibitem{} Hauk B., Mermilliod M., 1998, A\&AS, 129, 431
\bibitem{} Hernanz M., Jos\'e J., Coc A., Isern J., 1996, ApJ, 465, L27
\bibitem{} Iben I., Tutukov A. V., 1996, ApJS, 105, 145
\bibitem{} Johnson H. L., 1966, ARAA, 4, 193
\bibitem{} Mastrodemos N., Morris M., 1999, ApJ, 523, 357
\bibitem{} Odenwald S. F., 1986, ApJ, 307, 711
\bibitem{} Reyniers M., Van Winckel H., 2000, A\&A, in press,\\
astro-ph/0010486 (RVW)
\bibitem{} Robertson B. S. C., Feast M. W., 1981, MNRAS, 196, 111
\bibitem{} Starrfield S., Truran J. W., Sparks W. M., Arnould M., 1978, ApJ,
   222, 600
\bibitem{} Theuns T., Boffin H. M. J., Jorissen A., 1996, MNRAS, 280, 1264 
\bibitem{} Van Winckel H., 1999, in: Asymptotic Giant Branch Star, IAU
  Symp.\ 191, p.~465
\bibitem{} Waters L. B. F. M., Waelkens C., Mayor M., Trams N. R., 1993,
  A\&A, 269, 242 
\bibitem{} Whitelock P. A., 1987, PASP, 99, 573
\bibitem{} Whitelock P. A., 1988, in: The Symbiotic Phenomenon, (eds.)
 J. Mikolajewska et al., Kluwer, Dordrecht, p.~47 
\bibitem{} Whitelock P. A., Feast M. W., 2000, MNRAS, 319, 759
\bibitem{} Whitelock P. A., Menzies J. W., Feast M. W., Marang F., Carter B.,
   Roberts G., Catchpole R. M., Chapman J., 1994, MNRAS, 267, 711
\bibitem{} Whitelock P. A., Marang F., Feast M. W., 2000, MNRAS, 319, 728
   (WMF)
\end{thebibliography}
\end{document}